# Semi-autonomous, context-aware, agent using behaviour modelling and reputation systems to authorize data operation in the Internet of Things

Bertrand Copigneaux
Inno TSD
Sophia-Antipolis, FRANCE

*[1][2]Abstract*—In this paper we address the issue of gathering the "informed consent" of an end user in the Internet of Things. We start by evaluating the legal importance and some of the problems linked with this notion of informed consent in the specific context of the Internet of Things. From this assessment we propose an approach based on a semi-autonomous, rule based agent that centralize all authorization decisions on the personal data of a user and that is able to take decision on his behalf. We complete this initial agent by integrating context-awareness, behavior modeling and community based reputation system in the algorithm of the agent. The resulting system is a "smart" application, the "privacy butler" that can handle data operations on behalf of the end-user while keeping the user in control. We finally discuss some of the potential problems and improvements of the system.

*Keywords—Informed consent; authorization, data operation; context-aware; behaviour modelling; reputation systems; agent*

I. INTRODUCTION

The "Internet of Things" (IoT) concept envisions a worldwide network linking not only traditional ICT devices but also offering communication and computation capabilities and potentially autonomous behaviors to any device or system. The number and variety of foreseen applications use cases [1] of the IoT and their potential impacts on society are considered by many as tremendous [2].

Privacy concerns have accompanied the development of the IoT since the early days of ubiquitous computing [3]. The potential security and privacy issues now raised by the IoT are numerous and complex (user informed consent, continuity and availability of services, contextualization of risk, profiling, ownership of data, management and captivity of data, applicable legislation and enforcement…) and have been presented in the literature [4] [5].

[1] This work was published in Internet of Things (WF-IoT), 2014 IEEE World Forum, 6-8 March 2014, Seoul, P411-416, DOI: 10.1109/WF-IoT.2014.6803201, INSPEC: 14255656
[2] This work is currently supported by the BUTLER Project co-financed under the 7th framework program of the European Commission.

In this paper we focus on authorization of data operations in the IoT and propose an approach to ensuring the "informed consent" of the end user. The objective being to maximize the end user control and understanding on his data operations while minimizing the necessary number of operations by the end user.

As presented in section III, several approaches have been proposed in the existing state of the art. The solution we propose here, builds upon these existing approaches and combine them. Our analysis of this state of the art lead us to the conclusion that an automated approach is necessary to fit the vast number and complexity of data operation authorization decisions characteristic of a fully deployed IoT. The resulting system takes into account various sources (user defined rules, context, user behavior, community) to take decisions and advise the user on how to best protect his privacy by authorizing access to his data.

II. PROBLEM DEFINITION : INFORMED CONSENT

Informed consent is a term which originates in the medical research community and describes the fact that a person – such as a patient or a participant in a research study – has been fully informed about the benefits and risks of a medical procedure and has agreed on the medical procedure being undertaken on them. An informed consent can be said to have been given based upon a clear appreciation and understanding of the facts, implications, and future consequences of an action. In order to give informed consent, the individual concerned must have adequate reasoning faculties and be in possession of all relevant facts at the time consent is given.

From a legal perspective, the notion of consent is essential in data protection as the consent of a data subject is often necessary for a third party to legitimately process personal data. Within the European Union, the data protection directive [6] that defines conditions under which personal data can be processed specifies that the consent must be "freely given, specific and informed" and "unambiguous". The foreseen evolutions of this regulation [7] further strengthen this definition of consent by narrowing it to "explicit, clear affirmative action" excluding the possibility of implicit content.

Ensuring this level of "informed consent" can already be an issue in itself for traditional ICT applications, the technical and legal complexity of the problem being already an obstacle to informing potential end users. This has lead to the development of End User License Agreements (EULA) which are often too complex or too generic for most of the End User.

As a result a "consent fatigue" has developed, most end users accepting by default the license agreements and often without reading it. This is reinforced by the fact that the consequences of a potential privacy breach are distant and vague while the consequences of not accepting the end user license agreement are immediate and obvious (no access to the service or application). This effect has been observed and documented in [8] and [9].

This "informed consent" issue is further complicated by some the technical specificities of the Internet of Things. The tininess of the potential IoT devices, their distributed nature and integration into everyday life object complicate the necessity of information of the end users. The numbers of potential data operations in a fully deployed IoT [2] make even less practical than with the internet a systematic control of the data subject on each data operation.

The size of the data sets and complexity of the data operations taking place in cloud based infrastructure also enables advanced profiling which can reconstruct data and identify an end-user based on information that taken separately are not considered critical [10]. This ability is especially important as from the current [6] and foreseen [7] European legislation the legal definition of "personal data" is anything that can enable directly or indirectly (through profiling) the identification of the data subject.

Finally the distributed nature of the Internet of Things further complicates the informed consent issue as the end-user and data subject roles are more often separated than in traditional ICT applications: the distributed and decentralized nature of the IoT being therefore in conflict with the user centric problems of consent and privacy.

III. STATE OF THE ART

In the following section we discuss some of the techniques that have been proposed to address the informed consent and authorization problem and that have inspired our work.

A. *Standardized, human and machine readable license agreements*

The use of standardized license agreement, made available in three different but coherent version: full legal text (enforceable legal text), simplified text (short and understandable by most user), and machine readable version (enabling automated processing) has been experimented with some success in the copyright domain [11].

We propose in our setting to follow this approach to provide easily understandable and processable legal basis for authorizing access and use of data. It has the advantage to facilitate the information of the end user. However in itself, it doesn't bring solutions to the problem raised by the vast number of data authorization operations characteristic of a fully deployed IoT.

B. *Dynamic and context aware approach*

To face the dynamic and distributed nature of the IoT, it has been proposed to take in account the context in which an application operates to influence authorization [12] and disclosure of information [13].

The approach proposed in this paper is clearly based on these existing trends toward dynamic and context aware disclosure of information. However, as explained further on, we propose to complement this approach by adding the automated aspects of an agent based solution, the "Smartness" of behavior modeling and the social aspects of reputation systems.

C. *Semi-autonomous agent*

The use of a user centric, semi-autonomous agent to negotiate the consent of the user with third party applications based on user defined rules and preferences has been discussed in [14].

This approach has the advantage of keeping the end user in control (through the definition of rules) while allowing for scaling up to a large amount of data authorization operations (matching the needs of a fully deployed IoT). However the scope of the privacy coach proposed in [14] was limited to a single technology (RFID) and didn't take into account any context information.

D. *Behaviour modelling*

Advanced techniques have been developed to model and analyze the behavior of humans and their interactions with each other's and ICT technologies as presented in [15] [16] and [17]. These models can be used to create user adaptive systems [18] which provide services tuned to the individual preferences of the users.

Although this profiling can in itself lead to privacy issues, we argue that it could, with the necessary safeguards [19], be used to better understand individual user privacy requirements. We propose here to use profiling to automatically propose data operation authorization decisions to the user that match his previous decisions.

E. *Reputation systems*

A complementary and promising approach to further increase end user information on privacy and data protection in IoT would be to rely on reputation systems. Initial examples [20] and [21] of reputation based systems for trustworthy communications exists.

We propose to extend and generalize them to both become visible to end users and integrate user feedbacks and perceptions on IoT applications respect for privacy. Such ranking could not only provide information for end user but also be taken into account in semi-automatic selection of which node are trustworthy for information sharing.

## IV. OVERVIEW OF THE SYSTEM

The "privacy butler" system that we propose in this paper builds upon and combines these different trends to offer a solution to the informed consent problem in the Internet of Things.

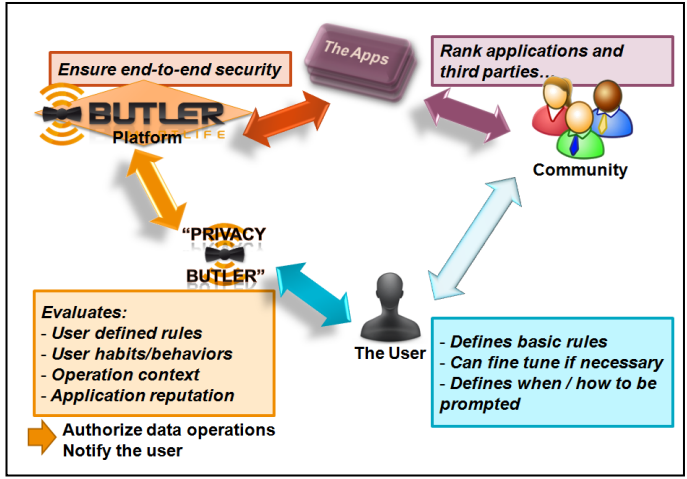

Fig. 1 - Overall System Description

As described in figure 1 the system is user centric. A graphic user interface enables the user to define a set of rules that should be both simple to comprehend for the user and complex enough to enable advanced users to fine tune if necessary. The user can also define how and when to be contacted by the system.

The privacy butler in itself is a semi-autonomous agent whose main role is to authorize or deny data operations on behalf of the user. To take each decision the agent evaluates the rules defined by the user but also context elements, and eventually user behavior and reputations of third parties.

To handle the reputation system, the user is able to participate to communities which evaluates and rank IoT applications and third parties (service providers, application developers…).

To ensure the end-to-end security of the interactions between the different elements of the systems and to ensure that third parties applications respect the decisions of the privacy butler, the whole system is built upon an IoT platform: the BUTLER platform. This platform has been defined in [22] and is currently in development. It is taken as a prerequisite of the system. Some architectural components of the BUTLER platform (behavior capture, data manager, context analysis…) are used by the "privacy butler" system.

## V. PRIVACY BUTLER ALGORITHM

In the following section we describe in more details the techniques and algorithms behind the "privacy butler"

### A. Semi autonomous, context-aware agent

The first difficulty to address in the set up of a "privacy butler" system is the definition of rules by the end user to handle his preferences. The main challenge is to offer a mechanism that is complex enough to address the diversity of IoT situation while remaining simple to use for the user. The approach we propose is to initially characterize data operations by three factors: the type of data to be processed, the action that will be performed on the data, and the third party that will perform the operation. Thus to a given data operation o a corresponding rule can be defined:

$$R(o) = A(o) + D(o) + P(o)$$

Where :

**A(o)** represents the type of action to be performed on the data.such as : *collecting data, keeping an history of collected data, profiling data, transferring data to a third party*. This type of operations are defined by the application provider

**D(o)** represents the type of data to be processed by data operation. To ease the definition of the rules groups can be defined by the end user such as: *Location data, Context Data, Direct Personal Information…*

**P(o)** represents the third party that will perform the operation. Here again the end user can regroup applications and organizations and other third parties in predefined or user defined groups, such as: *Friends, Coworkers, Highly Trusted Providers, Low Trust…*

Each rule can have three values: **Deny**, which strictly denies the data operation, **Allow**, which authorize the data operation or **Prompt**, which prompt the end user each time the data operation is requested. Examples of such rules would be:

*Allow - Collection operations, of Temperature data, by Bob.*
*Deny – keeping an History, of Location data, by Employer.*

Incomplete rules can be defined to enable the user to create more general rules:

*Deny - All operations, on Direct Personal Data data, by Low Trust group.*

In our example, when several rules can apply to a single operation, the priority is always on the Deny operation (but other behavior could be envisioned).

To take in account the context of the data operation, this initial model of rule definition is completed by the ability of the user to define context rules in the "privacy butler". In our example, and although more complete definition of context could be envisioned, we propose to follow the definition of context given in [13] where for a given scenario n the context is defined by the following statement:

$$Cn = F(N)+F(L)+F(P)+F(R)$$

Where:

**F(N)** represents the contextual parameter related to the network settings.

**F(L)** represents the current location of the object

**F(P)** includes the time and date of the interaction

**F(R)** represents the contextual parameters which identify an object from another (object identifier or IP address).

The user can thus define the context of a data operation (incomplete definition being possible) and associate it with a rule. The complete statement describing a rule thus becomes:

$$R(o) = A(o) + D(o) + P(o) + C(o)$$

Where:

**C(o)** represents the context of the data operation.

Figure 2 presents the workflow of operation to allow or deny data operations through this first version of the privacy manager. A context analysis is first made, then the "privacy butler" looks for an existing rule matching the situation and either applies it if it exists or asks the user how to proceed and if this decision should be made into a rule.

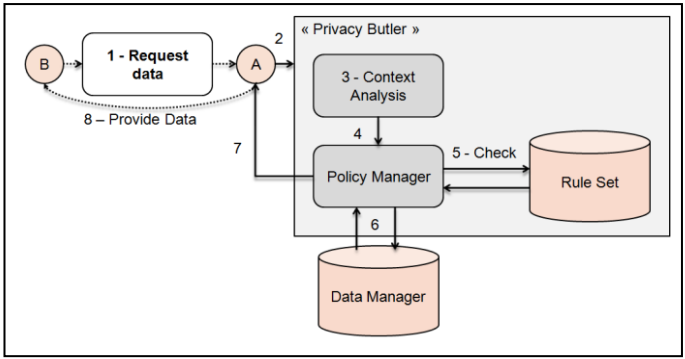

Fig. 2 – Simplified workflow

### B. Behaviour modelling and community based reputation

Having access to the user defined rules and to each individual decision made by the end user on how to handle his privacy, the "privacy butler" agent is in a privileged position to analyze the behavior of the end user.

Thus we propose to use the behavior modeling elements of the BUTLER IoT platform to analyze the individual decisions made by the user. By gathering knowledge on the decision, the system is eventually able to propose decisions and new rules to the user.

To further complement this decision engine, we propose to enable end users to share, on a voluntary basis and if necessary anonymously, some of their privacy settings with a community of users.

By gathering individual evaluations of applications, third parties and context situations the community system can be used by the agent to get an evaluation of a given situation.

In the advanced version of the "privacy butler" agent algorithm we therefore integrate behavior modeling and community system components.

Figure 3 presents the complete workflow of the system. The basic behavior is still in place (as described above in Figure 2, simplified workflow) and this workflow (steps 1, 2, 3, 4, 5, 9 10, 11) is followed when rules do exist in the Rule Set component. However when confronted with a situation not taken into account by a user defined rule the "privacy butler" examines, first the behavior modeling component and then the community system to gather insights on a potential decision to propose to the end user.

The behavior capture component is informed of every individual decision made by the user to authorize or deny access to his data and has access to the full set of rules defined in the Rule Set. Based on these data the behavior capture component infers (step 6) possible future decisions of the user based on the type of data (D(o)), type of data operation (A(o)), third parties involved (P(o)), context (C(o)).

If no clear rule proposition emerge from this analysis, the community reputation system steps in (step 7) and perform the same type of analysis on the community data. The community reputation systems characterize end users by a profile based on their defined rules set and previous decisions and can therefore advise them on what other users that have defined the same kind of privacy settings would do in a specific situation.

When a rule proposition has been defined, the Policy Manager validates the rule with the end user. The end user is given the opportunity to either enforce this decision only once or to define it as a rule in his Rule Set for all future actions matching this signature (type of data (D(o)), type of data operation (A(o)), third parties involved (P(o)), context (C(o)).

Table 1 presents a pseudo code version of the full algorithm for data operation evaluation by the "privacy butler" agent.

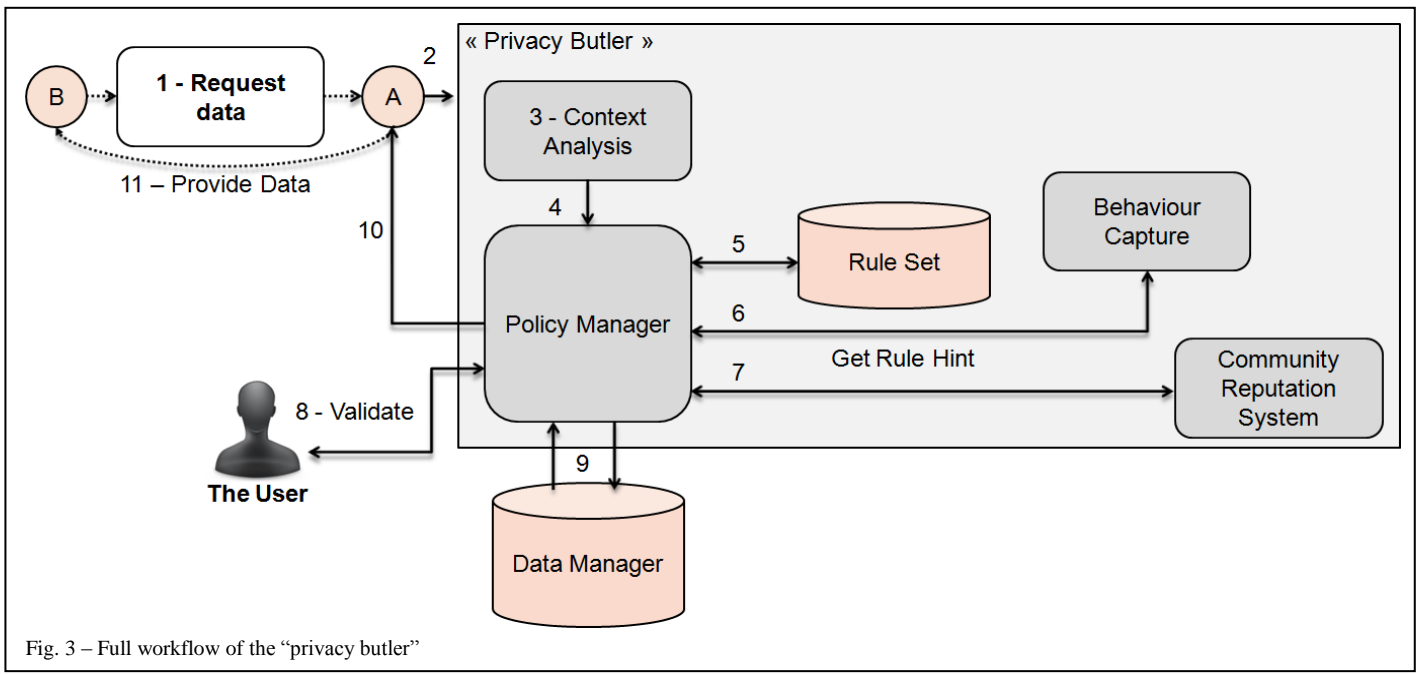

Fig. 3 – Full workflow of the "privacy butler"

```
//Privacy Manager receive getData request on data from third
party p willing to perform an action
ctxt = contextManager.getContext(current)
rule = ruleSet.getRule(p,data, action,ctxt)

if (rule != nil)
    applyRule(rule)
    behaviourModelling.registerAction(rule)

else if (rule == nil)
    ruleHint = behaviourModelling.getRuleHint(p, data, action,ctxt)

    if (ruleHint != nil)
        userInterface.promptUserAgree( ruleHint)
        userInterface.makeRule?(agree, ruleHint)
        behaviourModelling.registerAction(ruleHint)

    else if(ruleHint == nil)
        ruleHint = communityServer.getRuleHint(p, data, action,ctxt)

        if(ruleHint != nil)
            userInterface.promptUserAgree?(ruleHint)
            userInterface.makeRule?(ruleHint)
            behaviourModelling.registerAction(ruleHint)

        else if(ruleHint == nil)
            userInterface.promptUserAgree?(ruleHint)
            behaviourModelling.registerAction(ruleHint)
```

Table. 1 – "privacy butler" algorithm

## VI. DISCUSSIONS AND CONCLUSION

In this paper we introduced a novel approach to handle the authorization of data operation in the Internet of Things, combining different approaches previously introduced: a semi autonomous, rule based agent, which integrates context-awareness, behavior modeling and community based reputation systems.

We believe that such an approach can significantly improve the way the informed consent question is handled. The definition of rule can be very specific (taking into account the type of data, type of operation, identity of the third party requesting the data operation and context of the operation) enabling detailed control by the end user.

At the same time, the possibility to group elements together and the semi autonomous nature of the agent enable to limit the time needed for the end user to define his rules. The advices given by the behavior modeling system and the community based reputation system further simplify the task of the end user. As proposed in [14] the system could be initially populated by rules defined through a user-friendly questionnaire that introduce the user to the issue of privacy in the IoT.

One of the potential issues to be addressed is the "all or nothing" behavior of the "privacy butler" decisions. It can be expected that many third party applications will simply stop to function altogether if a data operation they requested is denied by the privacy manager. This issue could be addressed by introducing different level of data obfuscation as alternative to the deny operation. Examples of this mechanism are presented in detail in [13] for location data but could probably be generalized to other type of data.

Another potential issue to be further addressed is the increasing differentiation in the IoT between the end user and the data subject (i.e. an IoT application, such as a sensor network, may collect personal data on data subjects that are not directly users and therefore may not be able to give consent). We believe that this issue, as well as the correct enforcement by the IoT applications of the rules defined by the "privacy butler" could be handled by the adoption of common, standardized IoT platforms, with end-to-end security (such as the one the BUTLER project aims to develop) and supported by specific governance.

ACKNOWLEDGMENT

Special thanks to Dr. Judicaël Ribault for his careful review, and insightful comments.